\documentclass[twocolumn,showpacs,preprintnumbers,amsmath,amssymb,prl]{revtex4}
\usepackage{graphicx}
\usepackage{dcolumn}
\usepackage{bm}
\begin{document}

\title{Coexisting pseudogap, charge transfer gap, and Mott gap energy scales in the
resonant inelastic x-ray scattering spectra of electron-doped cuprates}

\author{Susmita Basak$^1$, Tanmoy Das$^1$, Hsin
Lin$^1$, M.Z. Hasan$^2$, R.S. Markiewicz$^1$, and A. Bansil$^1$}

\address{1: Physics Department, Northeastern
University, Boston MA
02115, USA}

\address{2: Department of Physics, Joseph Henry Laboratories of Physics, Princeton
University, Princeton NJ 08544, USA}
\date{\today}

\begin{abstract}

We present a computation of Cu K-edge resonant inelastic x-ray scattering
(RIXS) spectra for electron-doped cuprates which includes coupling to
bosonic fluctuations.  Comparison with experiment over a wide range of
energy and momentum transfers allows us to identify the signatures of three key normal-state
energy scales: the pseudogap, charge transfer gap, and Mott gap.  The
calculations involve a three band Hubbard Hamiltonian based on Cu
$d_{x^2-y^2}$ and O $p_x$,~$p_y$ orbitals, with a self-energy correction which arises due to spin
and charge fluctuations. Our theory reproduces characteristic features e.g., gap collapse, large spectral weight broadening, and spectral weight transfer as a function of doping, as seen in experiments.
\end{abstract}

\maketitle\narrowtext


Cuprates are widely believed to be charge-transfer insulators\cite{ZSA}, with a Mott gap between
Cu-$d_{x^2-y^2}$ orbitals much larger than the charge transfer gap between
Cu-$d$ and O-$p$ orbitals.  The upper (UHB) and lower Hubbard band (LHB)
of the Cu orbitals are intimately related to the antibonding and bonding bands of the
three band model, and it is important to understand how the strong
correlations of Mott physics modify these bands from the conventional
LDA-based picture.  Meanwhile, a third energy scale, the pseudogap scale,
has been found experimentally, and its origin and relation to the other
two scales continues to be a matter of intense debate.  Here we model electron-doped
$\rm Nd_{2-x}Ce_xCuO_{4\pm\delta}$ (NCCO), for which the pseudogap is well described as
a competing antiferromagnetic (AF) order.

\begin{figure}[htp]
\centering
\hskip-0.5 cm
\rotatebox{0}{\scalebox{0.16}{\includegraphics{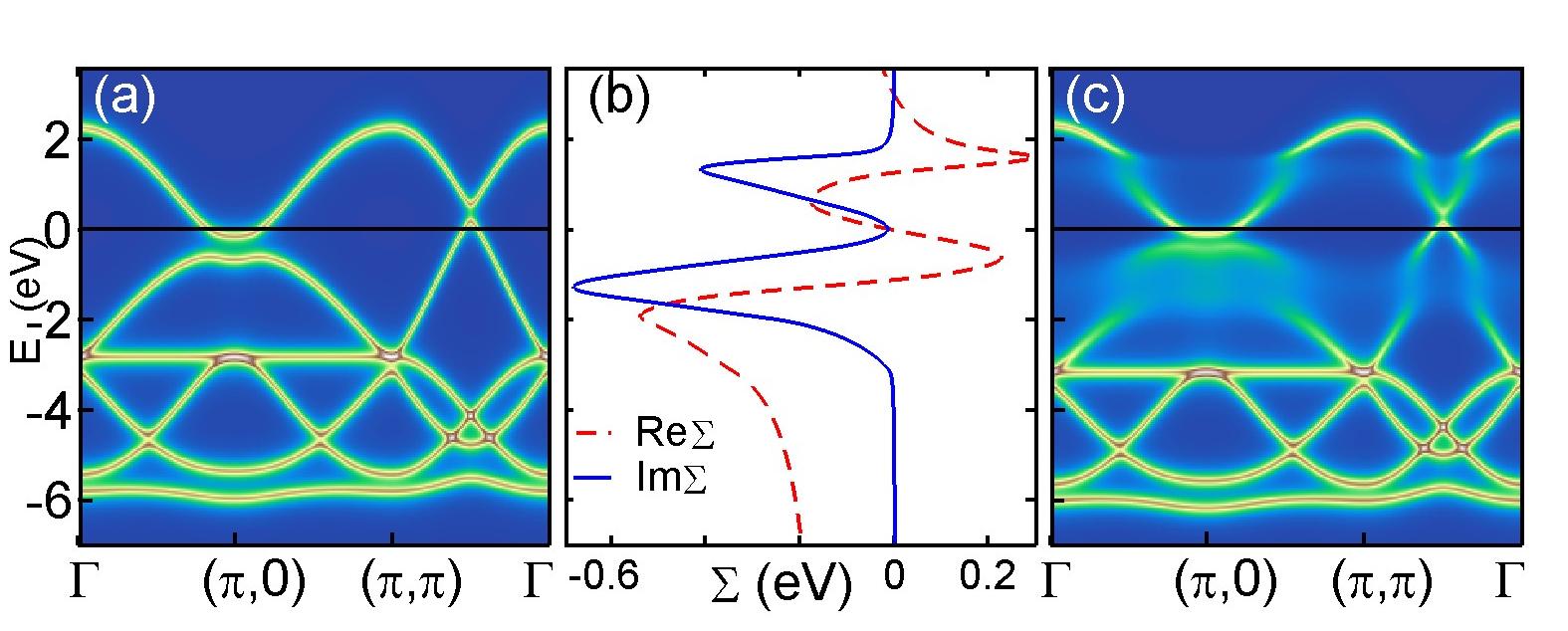}}}
\caption{(Color online) (a) Spectral weight of electronic states in NCCO for $x=0.14$.
A constant imaginary part of magnitude $0.5$ eV is added to broaden
the spectra.
(b) {Calculated self energy for anti-bonding bands. The red dashed curve and the blue solid curve are respectively
real and imaginary parts of the self-energy.}
(c)  Spectral weight as in (a), but
modified by the self energy of (b).}
\label{AnSEx15}
\end{figure}

Experimental access to the LHB and/or the bonding band has proven difficult and the corresponding optical interband transitions have not been observed.  Moreover, while the antibonding
$d_{x^2-y^2}$ band lies at the top of the $d$-bands, the bonding
$d_{x^2-y^2}$ band is found in LDA to lie at the bottom of a veritable
`spaghetti' of $d$-bands and their associated oxygen orbitals, nearly $6$~eV
below the Fermi level as seen in Fig. \ref{AnSEx15}. Therefore, it is difficult to
extract this band from ARPES data. On the other hand, RIXS is a local
probe directly rearranging the Cu and O orbitals, and as such can
provide selective access to the bonding bands. Indeed, RIXS
experiments report a strong feature in most cuprates in the $6$-$8$ eV range
which has been associated with this band \cite{x0higherEcuts,Hasan2,Kim,Ishii2,Hasan3,simon}.
In this article we show that by incorporating
strong renormalization of the near Fermi energy bands by magnon
fluctuations \cite{water3,tanmoysw}, the high-energy RIXS features are
indeed consistent with transitions from the LHB to the UHB.
This resolves a puzzling discrepancy in earlier calculations \cite{bobrixs}
which were unable to fit both the low and high energy parts of the spectrum. We also capture
another important feature of the spectrum, the realistic broadening, which arises due to the
strong coupling to bosonic quasiparticles.


Remarkably, we find that all three energy scales are strongly influenced by the Hubbard $U$.
The three energy scales are the following: 1) the Mott gap scale which is the result of transitions
from LHB to UHB, 2) the charge transfer gap scale which persists as a residual feature into the overdoped regime
and 3) the pseudogap or AF gap scale which collapses in a quantum critical point near optimal doping.  For convenience,
we label the AF-split subbands of the antibonding bands as the lower (LMB) and upper ( UMB) magnetic bands.
Cuprate magnetism naturally separates into two regimes: at high energies
{\it Mott physics} produces localized spin singlets on each copper site,
splitting the Cu dispersion by an energy $\sim U$ into upper and lower Hubbard bands.
In the presence of hybridization with oxygens, the LHB [UHB] becomes
identified with the bonding [antibonding] band of the three-band model.
At lower energies, these singlets interact on different sites, leading to
magnetic gaps in both UHB and LHB of magnitude $\sim m_dU$ via more conventional {\it Slater
physics} associated with AF order, where $m_d$ is the magnetization on Cu.
The Mott physics arises as an {\it emergent phenomenon}.
 When the AF gaps open at half filling, hybridization between Cu and O is mostly lost.  For instance, in the antibonding band
electrons in the UMB have mainly Cu character, while the opposite happens in the bonding band \cite{bobrixs}.
Consequently, the states near the top of the
lower magnetic band are of nearly pure oxygen character \cite{bobrixs}.Thus, due
to strong correlations, the `charge-transfer' gap at half filling coincides with the AF gap. At finite
doping, these two features separate in energy: the AF gap collapses rapidly \cite{foot1}, while a
residual charge-transfer gap persists in optical spectra at high energies, due to strong
magnetic fluctuations, closely related to the high energy kink (HEK), or
`waterfall' effect seen in photoemission \cite{water3,basak}. Here we show
that this residual charge-transfer gap is also present in RIXS.

In K-edge RIXS the incident x-ray excites a Cu
$1s\rightarrow 4p$ transition with an intermediate state shakeup involving
mainly Cu $d_{x^2-y^2}$ and O $p$ states. Within the RPA framework, the RIXS cross section
for this process is \cite{Igarashi,bobrixs,LiHasan}
\begin{eqnarray}\label{Wqw}
W({\bf q},\omega,\omega_i) &=& (2\pi)^3N|w(\omega,\omega_i)|^2
\nonumber\\
&\sum_{\mu}&{\rm Im}\left[Y^{+-}_{\mu,\mu} ({\bf q},\omega)\right]
|\alpha_{\mu}|^2 \cos{\big(2{\bf q}\cdot{\bf R}_{\mu}\big)}
\end{eqnarray}
Here $\omega_i$ is the initial photon
energy (taken to be -5 eV) and $\omega$, $\textbf{q}$ are the energy and the momentum,
respectively, which are transferred in the RIXS process. $w(\omega,\omega_i)$ contains all the matrix-element information
of the initial and final state transition probabilities \cite{bobrixs}, $N$ is the total number of Cu atoms and ${\bf R}_{\mu}$ is
the position of the $\mu^{\rm th}$ orbital present in the intermediate
state. The nearest neighbor (NN) O excitations and second NN
Cu excitations are included via $\alpha_1$ and $\alpha_2$, respectively. We assume small
values of $\alpha_1=0.1$, $\alpha_2=0.05$ in this study, whereas $\alpha_0$ is equal to 1.
The transferred momentum and energy are then shared
by the electron-hole pair created in the intermediate state from Cu
$d_{x^2-y^2}$ and O $p$ bands, $Y^{+-}_{\mu,\mu} ({\bf
q},\omega)$. In the Keldysh formalism it takes the form of a charge
correlation function or the joint density of states (JDOS) (in
real time) as $Y^{+-}_{\mu'\sigma',\mu\sigma}(q,t'-t) =
\langle \rho_{\bf{q},\mu' \sigma'}(t')\rho_{-\bf{q},\mu
\sigma}(t)\rangle$ with $\rho_{\bf{q},\mu \sigma}(t)$ representing
the charge operator. It is straightforward to show that $Y$ can
be calculated as the convolution between the spectral weights ($A$)
of the filled and empty states \cite{Igarashi,allen} as
\begin{eqnarray}\label{Y}
Y^{+-}_{\mu'\sigma',\mu\sigma}({\bf q},\omega) &=&
\sum^{\prime}_{k}\int d\omega_1\int d\omega_2
A_{\mu\mu'}(k+q,\omega_1)
\nonumber\\
&\times&A_{\mu'\mu}(k,\omega_2)\frac{f(\omega_2)-f(\omega_1)}{\omega+i\delta+\omega_2-\omega_1}
\end{eqnarray}
where $f(\omega)$ is Fermi function and $\sigma$ is the spin
index. The prime in the $k-$summation means that the summation is
restricted to the AF zone.

RIXS spectra are calculated using Eq.~\ref{Y} in which the spectral weight
$A$ is computed using a {three-band Hubbard model with the Hamiltonian:}
\begin{eqnarray}\label{Hamil}
H&=&\sum_j (\Delta_{d0}d_j^{\dagger}d_j + Un_{j\uparrow}n_{j\downarrow}) + \sum_i
U_pn_{i\uparrow}n_{i\downarrow}
\nonumber\\
&+&\sum_{<i,j>} t_{CuO}(d_j^{\dagger}p_i+ (c.c)) + \sum_{<i,i^{\prime}>}
t_{OO}(p_i^{\dagger}p_i^{\prime}+ (c.c)),
\end{eqnarray}
{where $\Delta_{d0}$ is the (bare) difference between the onsite
energy levels of Cu $d_{x^2-y^2}$ and O $p-\sigma$, $t_{CuO}$ the copper-d
oxygen-p hopping parameter, $t_{OO}$ the oxygen-oxygen hopping parameter and $U$ ($U_p$) the
Hubbard interaction parameter on Cu (O). $n_{j} = d_j^{\dagger}d_j$ and $n_{i} =
p_i^{\dagger}p_i$ are the number operators for Cu-d and O-p electrons, respectively. The
equations were solved at Hartree-Fock (HF) level to obtain a self-consistent mean-field solution.
Hartree corrections lead to a renormalized Cu-O splitting parameter $ \Delta =
\Delta_{d0}+ Un_d/2- U_pn_p/2,$ where $n_d$ ($n_p$) is the average electron density on Cu(O) \cite{foot2}.
The AF order splits the three bands into six bands as seen in  Fig.~\ref{AnSEx15}(a).  Since self-energy corrections are explicitly
included, we use bare LDA-like dispersions \cite{foot2} in the three-band
model rather than the dressed, experimental dispersions \cite{bobrixs}.  Thus hopping parameters are
taken from LDA while interaction parameters $U$ and $\Delta_{d0}$ are adjusted to optimize agreement
between the antibonding band splitting and earlier one band results \cite{bobrixs,foot2,footAB2,footAB3,OKA}.
When this is done, we find that $\Delta_{d0}$ is small and negative while $U$ has a very weak doping
dependence \cite{foot3}.

The renormalization of the antibonding band due to bosonic fluctuations is calculated via a
self-energy based on the QP-GW formalism \cite{SEtd},
%
\begin{eqnarray}\label{Aselfeng}
\Sigma({\bf k},i\omega_n)=
\hspace{5cm}\nonumber\\
\sum_{\bf q}
\int_{-\infty}^{\infty}\frac{d\omega^{\prime}}{2\pi}\Gamma
G({\bf k}-{\bf q},i\omega_n+\omega^{\prime}) W({\bf
q},\omega^{\prime}).
\end{eqnarray}
Here $\Gamma$ is the vertex correction and $W=U^2\chi$ is the
interaction term which includes both spin and charge fluctuations.
AF order in included along the lines of Ref.~\onlinecite{basak}
where the effective AF gap is kept the same as in the one
band model.  Finally the self-energy ($\Sigma$) is incorporated
 into the three band dispersion via Dyson's equation
$G^{-1}=G_0^{-1}-\Sigma$ and $A$ is computed from the dressed $G$.
Our calculation includes only fluctuations associated with
the band closest to the Fermi level, which produces negligible
broadening for $\omega > 4$ eV. Therefore for higher energy bands we include
a constant broadening, $\Sigma^{''} = 0.5$ eV, consistent with the ARPES data \cite{lanzara}.

Figure~\ref{AnSEx15} shows how self energy effects modify the
dispersion of the various bands of the three-band AF model for $x=0.14$, comparing bare (a) and dressed (c) bands.  The
imaginary part of the self-energy, plotted in Fig.~\ref{AnSEx15}(b),
attains a maximum around $1.7$ to $2$ eV, which leads to a strong
broadening of the spectral weight in this energy range, both below and
above the Fermi level (denoted by the black line), producing a
characteristic kink or `waterfall' effect in the dispersion.
We will see in connection with Fig.~\ref{figSE} below that
this `waterfall' effect leads
to a significant broadening in the RIXS spectrum since the spectrum of
Eq.~\ref{Y} involves a convolution of the filled and empty states.
Fig.~\ref{AnSEx15}(c) also shows that the self-energy softens the low
energy bands nearest the Fermi level. This
renormalization should also show up in the lowest branch of the RIXS spectrum,
but this is restricted to very low energies and does not appear prominently
in Fig.~\ref{figSE}.


Figure~\ref{figSE} shows the calculated RIXS spectra of
NCCO for $x=0$ and $x=0.14$, reflecting the modulation of the spectral intensities of Fig. 1 via matrix element effects, which are well known to be important in various highly resolved spectroscopies.\cite{footAB5,footAB7,footAB8} Frames (a) and (d) include AF order but without self energy corrections, whereas the
calculations in frames (b) and (e) include the self energy.
The high intensities at energies around $5.6$ eV involve the
transition from the lower Hubbard band to the unoccupied states of the
antibonding band, reflecting the Mott gap feature. This `6~eV' feature is present for all dopings. At half filling, in frames (a) and (b), the high
intensities around $2$ eV occur due to the transition within the
antibonding Cu-O band across the AF gap. This gap collapses with doping
and as a result we find a smaller AF gap at $14 \%$ electron
doping in frames (d) and (e), close to the QCP, consistent with earlier
results \cite{LiHasan}.  A key result is that the self energy produces a realistic broadening comparable to that observed experimentally.

\begin{figure}[htp]
\centering
\hskip-0.7 cm
\rotatebox{270}{\scalebox{0.15}{\includegraphics{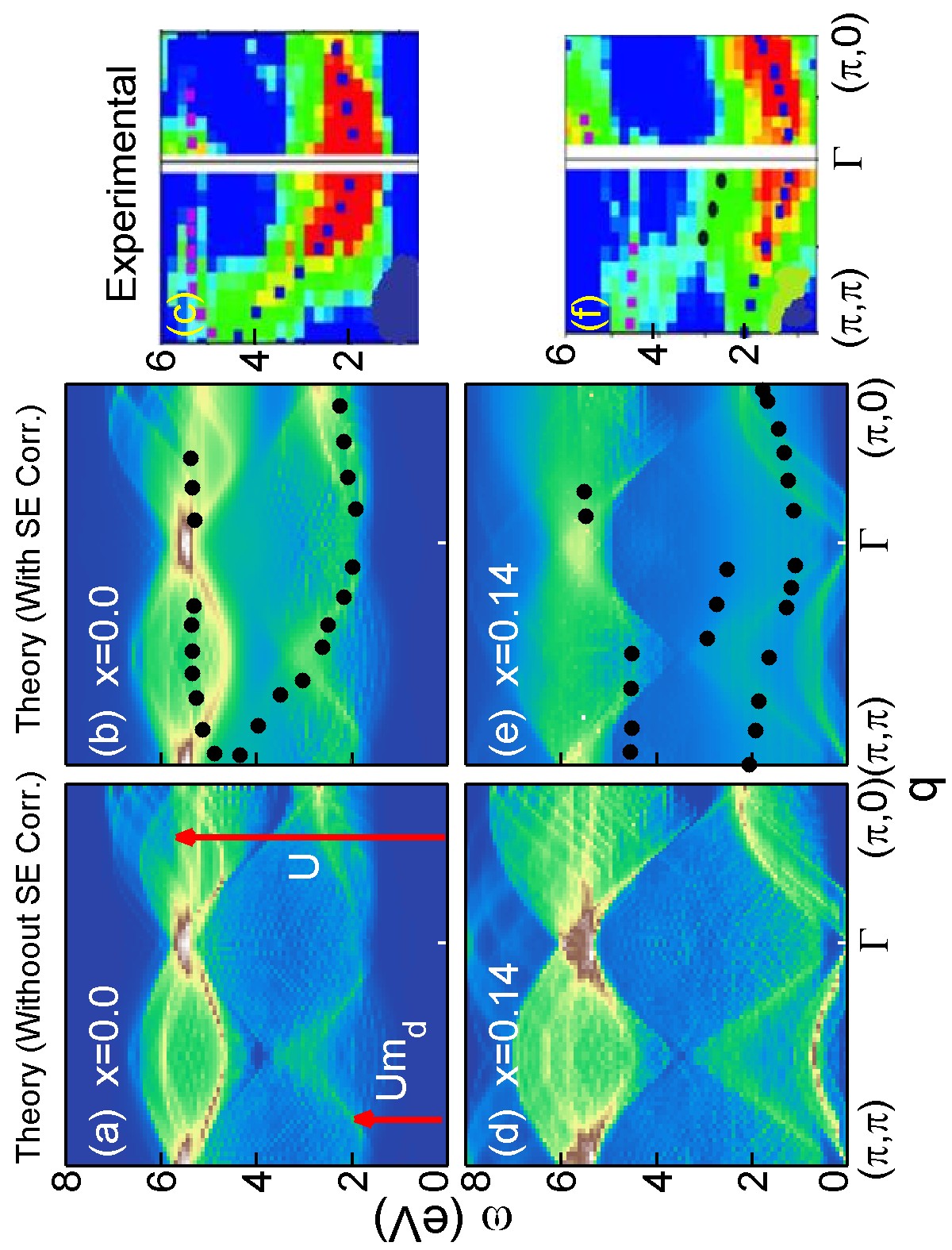}}}
\caption{(Color online)  RIXS spectra from NCCO for $x=0$: (a) theory without and (b)
with self energy corrections, and (c) experiment
\cite{LiHasan}.  (d)-(f) Similar figures for $x=0.14$.}
\label{figSE}
\end{figure}

In the RIXS calculations, the 6~eV feature is the most intense in the spectrum,
consistent with most early experiments on a variety of cuprates \cite{hill,Abbamonte,x0higherEcuts,Hasan,Ishii,Lu},
but more recent experiments\cite{Hill2}, including those of Figs.~2(c,f)\cite{Li}, employ a range of $\omega_i$ where the 6~eV feature is suppressed and the lower energy features can be more easily probed.  Except for this feature, most features in the calculated RIXS intensities follow the experimental trends. In the
undoped cuprate in Fig. \ref{figSE} (b), we observe a broad peak at
$\Gamma$ around $2.5$ eV, with the intensity decreasing around the zone
corner $(\pi,\pi)$ while it remains strong around $(\pi,0)$.  A similar
level of agreement is found in the case of $14 \%$ electron doping in
Fig.~\ref{figSE} in panels (d)-(f) \cite{foot4}. The black dots in panels (b) and (e)
represent the peaks of the experimental spectra, reproducing the blue, black,
and purple dots of panels (c) and (f). The agreement is remarkable for both
dopings, $x=0$ in frames (b) and~(c) and $x=0.14$ in frames (e)
and~(f) \cite{foot8}. Results for $x$=0.09 are similar, and are omitted
for brevity.

We comment here on the three energy scales.  While the dispersions which follow
from Eq.~3 are rather complicated, we find numerically that the Mott gap is approximately
equal to $U$ and the AF gap to $Um_d$, as illustrated by arrows in Fig.~2(a).  Also, the
charge transfer energy is the difference between the average oxygen energy and the upper Cu band \cite{ZSA},
which we find to be $\sim U/2$.  Thus all three energy scales are controlled by $U$.
In our calculation the $6$ eV
feature represents transitions across the true Mott gap, and the good
agreement with experiment indicates that RIXS can be used to probe this
important scale and how it is modified by hybridization with oxygens -- is
the bonding band split as our calculations suggest?  This feature will be
discussed further below when we describe fits to individual $q$-cuts of
the RIXS spectra. In optical spectra at half filling the $\sim$2~eV charge transfer gap is indistinguishable
from the AF gap\cite{tanmoysw}.  At finite doping these two features
separate, with the AF gap reflected as a
midinfrared peak which collapses rapidly with doping, while a residual
charge transfer gap persists as a weak feature near $2$ eV in the strongly
doped regime.  A similar evolution is found in RIXS. The RIXS leading edge
follows the doping dependence of the AF gap \cite{bobrixs,LiHasan},
while in Fig.~3 we show that a residual charge transfer gap feature can be
seen in the RIXS spectra near the $\Gamma$ point.  Our three band calculation successfully reproduces the experimental finding that the magnetization scales with the AF gap \cite{Mang,MarkieAF}.
Our calculated
three-band RIXS spectrum modified by self-energy beautifully displays the
broad feature around $2$ eV at $\Gamma$ in panel (b), in good agreement
with experimental results in panel (a). Panels (c) and (d) display RIXS
intensities obtained from theory (blue) as well as experiment (red dots)
as a function of energy at $\Gamma$, compared with the optical
spectra \cite{onoseprb} (green dashes) for $x = 0.10$ and $x=0$,
respectively.  The peak of experimental intensity is shifted towards
slightly higher energy than the theoretical intensity, but the broadening
is comparable in both cases.

\begin{figure}[htp]
\centering
\rotatebox{270}{\scalebox{0.18}{\includegraphics{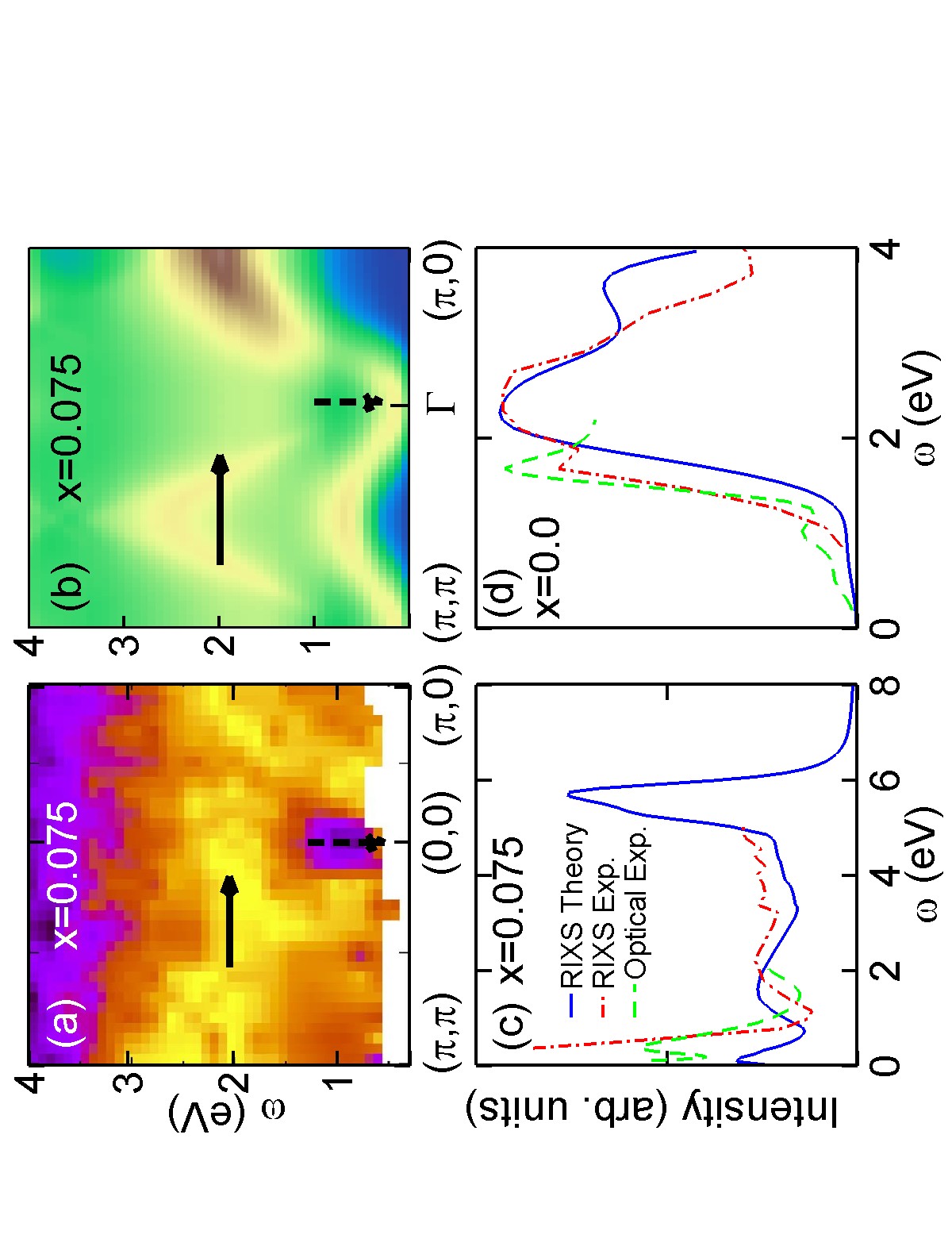}}}
\caption{{(Color online) RIXS spectra from NCCO for $x=0.075$, (a) experiment} \cite{Ishii} and {(b)
theory. Intensity cuts along $\Gamma$}, (c) for $x=0.075$ and (d) for
$x=0$ with RIXS theory (blue solid line), RIXS experiment (red dot-dashed line) and optical
experiment (green dashed line). In frames (a) and (b), solid arrows
indicate intensity peaks along $\Gamma$, while dashed arrows indicate cuts
taken in frame (c). In frames (c) and (d), the theoretical curves have
been convoluted with a Gaussian broadening of $200$~meV to mimic
experimental resolution.
}
\label{expishii}
\end{figure}
%

For more quantitative estimates of the broadening, Fig. \ref{cuts}
compares theoretical (blue solid) and experimental (red dashed line) RIXS intensities as a
function of $\omega$ for several constant $q$-cuts.  There is an overall
good agreement in peak positions as well as lineshapes and broadening for
all momenta.  In particular, panel (b) shows the high energy RIXS feature
from Ref.~\cite{x0higherEcuts}.  The good agreement with theory strongly
suggests the identification of this feature with the LHB in NCCO.  A
similar peak is found in all cuprates, as would be expected for Mott
physics.

\begin{figure}[htp]
\centering
\hskip-0.2 cm
\rotatebox{270}{\scalebox{0.18}{\includegraphics{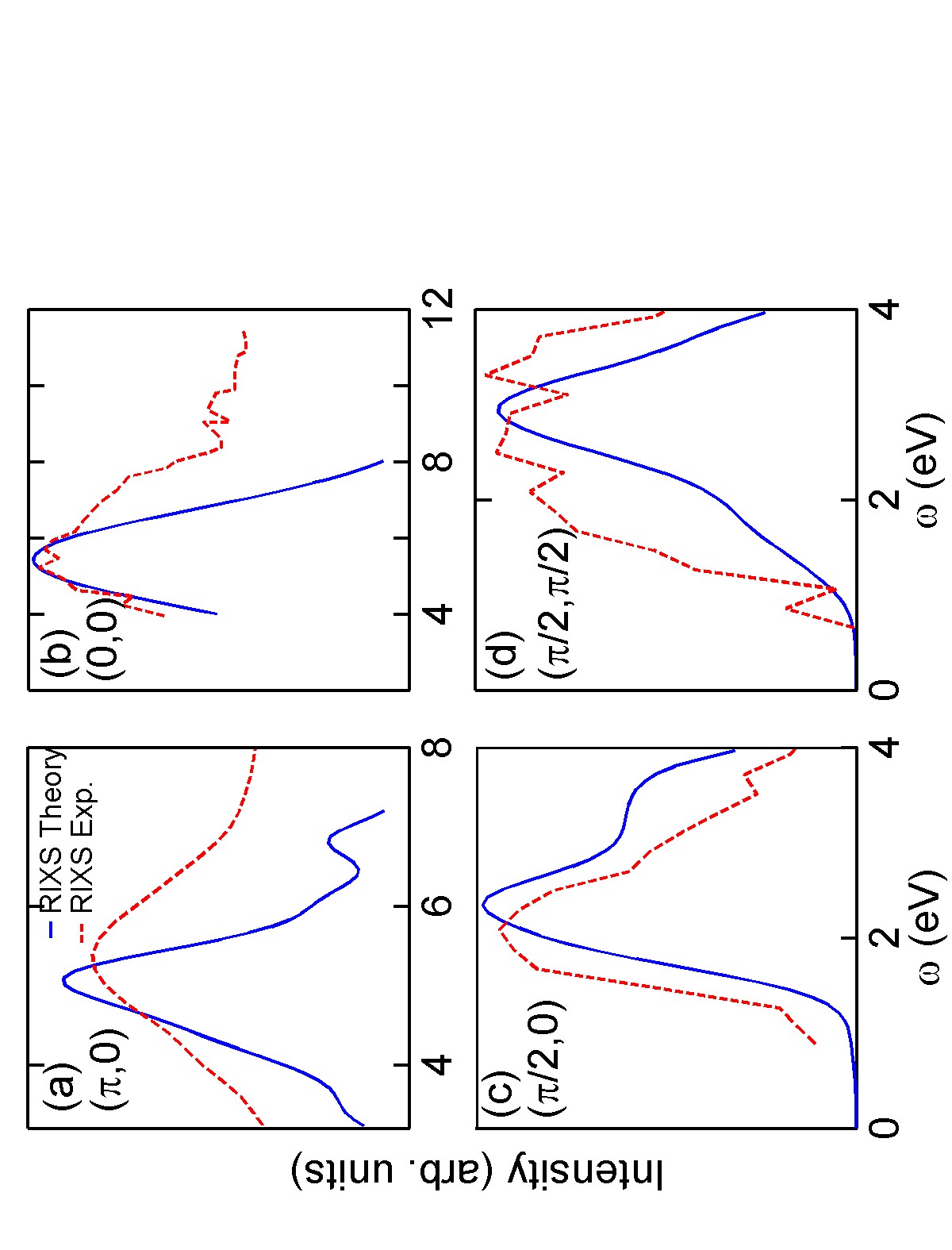}}}
\caption{(Color online) Comparison of theoretical (blue solid line) and experimental (red dashed line) spectra
at half-filling with $q$ fixed at (a): ($\pi,0$),~(b): $\Gamma$, (c):
($\pi/2,0$), (d): ($\pi/2,\pi/2$). Panel (b) displays the high-energy RIXS
peak from Ref.~\onlinecite{x0higherEcuts}.
The theoretical curves have been convoluted with a
Gaussian broadening of $200$~meV to mimic experimental resolution.}
\label{cuts}
\end{figure}

In conclusion, we find that RIXS is a suitable probe across all energy
scales, including AF gap, charge-transfer, and Mott physics.  We
provide a three-band model that is capable of explaining the experimental
RIXS spectra over the entire energy and doping range. We find a good
correspondence between the RIXS spectra at $\Gamma$ and the optical
spectrum, but RIXS has the additional advantage of full momentum-space
resolution. While we have concentrated on the electron doped cuprates, our
model should apply equally well to the hole doped case.

\begin{acknowledgments}

We thank A. Lanzara and A. Tremblay for very useful conversations.  This
work is supported by the U.S. Department of Energy, Office of Science,
Division of Materials Science and Engineering grants DE-FG02-07ER46352 
and DE-AC03-76SF00098, and benefited from the collaboration supported by the
Computational Materials Science Network (CMSN) program
under grant number DE-FG02-08ER46540, and from the allocation
of supercomputer time at NERSC and Northeastern University's Advanced
Scientific Computation Center (ASCC).

\end{acknowledgments}



\begin{thebibliography}{99}
\bibitem{foot1}For finite hole doping, the AF order can equally well be modeled by a competing magnetic, charge,
or flux phase order.
\bibitem{foot2}The values of the three-band model parameters used in our
calculations are as follows. $t_{CuO}=1.3$ eV, $t_{OO}=-0.65$ eV, 
$\Delta = -0.5,-0.38,-0.41$ eV and $m_d = 0.2, 0.12,0.06$ for $x = 0.0, 0.075, 0.14$, respectively. We have chosen $\Delta_{d0}= -0.755$ eV which is doping independent. The hopping parameters are obtained by using a tight-binding fitting to ab-initio results, and are taken to be doping independent assuming a rigid band picture\cite{footAB2}. It will be interesting to examine doping effects via first principles approaches \cite{footAB3}. The magnetization is found self-consistently.  The small value of $\Delta$ is consistent with Ref.~\onlinecite{OKA}.
\bibitem{foot3}We find that a smaller value of the effective Hubbard
parameter ($U=6.5$ eV instead of $7.2$ eV) leads to better agreement with experiments at
half-filling.  For 14\% doping, however, we use the same renormalized $U=5.7$~eV as Ref. \cite{bobrixs}.
\bibitem{foot8}
We note two slight disagreements: (1) along $\Gamma\rightarrow (\pi,\pi)$
the theoretical curves show a reflection symmetry about $(\pi/2,\pi/2)$
associated with long range AF order, which is weaker or absent in
experiment, suggestive of a fluctuating pseudogap.  (2) While the
high-energy feature near 5~eV, associated with transitions to the LHB, is
in good agreement with experiment in the undoped sample in
Figs.~\ref{figSE}
(a)-(c) and along $\Gamma\rightarrow (\pi,0)$ in the $14 \%$ sample in
Figs.~\ref{figSE} (d)-(f), experiments in the latter sample along
$\Gamma\rightarrow (\pi,\pi)$ show an anomalous softening not captured in
the present calculation.
\end{thebibliography}
\end{document}